\documentclass[10pt,twocolumn,letterpaper]{article}
\usepackage{graphicx} 
\usepackage{grffile}
\usepackage[table,xcdraw]{xcolor}
\usepackage[utf8]{inputenc}
\usepackage{algpseudocode}
\usepackage[pagenumbers]{cvpr} 

\definecolor{cvprblue}{rgb}{0.21,0.49,0.74}
\usepackage[pagebackref,breaklinks,colorlinks,citecolor=cvprblue]{hyperref}

\title{Simple 2D Convolutional Neural Network-based Approach\\ for COVID-19 Detection}

\author{$^\dagger$$^1$Chih-Chung Hsu, $^\ddagger$$^1$Chia-Ming Lee, $^2$Yang Fan Chiang,  $^1$Yi-Shiuan Chou, \\$^1$Chih-Yu Jiang,$^1$Shen-Chieh Tai, $^1$Chi-Han Tsai\\
$^1$Institute of Data Science, National Cheng Kung University, Taiwan\\
$^2$Department of Electrical Engineering, National Cheng Kung University, Taiwan\\
{\tt\small $^\dagger$cchsu@gs.ncku.edu.tw, $^\ddagger$zuw408421476@gmail.com}
}

\begin{document}
\maketitle
\begin{abstract}

%
This study explores the use of deep learning techniques for analyzing lung Computed Tomography (CT) images. Classic deep learning approaches face challenges with varying slice counts and resolutions in CT images, a diversity arising from the utilization of assorted scanning equipment. Typically, predictions are made on single slices which are then combined for a comprehensive outcome. Yet, this method does not incorporate learning features specific to each slice, leading to a compromise in effectiveness. To address these challenges, we propose an advanced Spatial-Slice Feature Learning (SSFL++) framework specifically tailored for CT scans. It aims to filter out out-of-distribution (OOD) data within the entire CT scan, allowing us to select essential spatial-slice features for analysis by reducing data redundancy by 70\%. Additionally, we introduce a Kernel-Density-based slice Sampling (KDS) method to enhance stability during training and inference phases, thereby accelerating convergence and enhancing overall performance. Remarkably, our experiments reveal that our model achieves promising results with a simple EfficientNet-2D (E2D) model. The effectiveness of our approach is confirmed on the COVID-19-CT-DB datasets provided by the DEF-AI-MIA workshop. 
\end{abstract}    
\section{Introduction}
\label{sec:intro}

Computed Tomography (CT) \cite{ct} has emerged as an indispensable tool for disease detection and management. Its capability to uncover abnormalities within the body, including ground-glass opacities and bilateral patchy shadows, is vital for the early detection and monitoring of diseases. In the context of diagnosing COVID-19, medical professionals rely heavily on the analysis of lung CT scans. However, considering that a single patient's CT scan may comprise hundreds of images, manual inspection becomes an exceedingly time-consuming task, particularly when physicians are required to assess CT scans from a significant number of patients. This situation could lead to the occurrence of false negative samples due to the overwhelming volume of data.

The swift progress in deep learning (DL) technologies has positioned DL methods \cite{Thyroid,Thyroid2,Thyroid3,Thyroid4,Thyroid5,ml,ml2} at the forefront for their capacity to rapidly and precisely detect features of COVID-19, efficiently processing vast amounts of data. Moreover, convolutional neural networks (CNNs) have demonstrated superior effectiveness compared to both frequency-domain-based \cite{frequency,frequency2} and low-level features-based approaches \cite{sift} for CT image analysis.

In response to the widespread outbreak of COVID-19, Kolliaz \emph{et al.} introduced the COVID-19-CT-DB dataset \cite{arsenos2022large,arsenos2023data,kollias2020deep,kollias2020transparent,kollias2022ai,kollias2023deep,kollias2021mia,kollias2023ai}, a comprehensive collection that includes a large volume of labeled data for both COVID-19 and non-COVID-19 cases. This initiative significantly advances deep learning (DL) methodologies, addressing the critical need for high-quality datasets essential for DL-based analysis in the fight against COVID-19. Numerous researchers have subsequently developed various methods aimed at enhancing COVID-19 detection tasks, contributing valuable strategies to the ongoing effort to manage the pandemic effectively \cite{chen2021adaptive,hsu2022spatiotemporal,hsu2023bag,zhang2021efficient}.

Although CT imaging is a powerful tool for detecting abnormalities, it faces challenges related to varying resolutions and qualities arising from different data servers and scanning machines. The resolution and number of slices in CT images can differ depending on the specific scanning equipment used, potentially requiring researchers to create complex network designs. Moreover, medical analysis for COVID-19, unlike standard DL-based tasks that prioritize performance and application, necessitates maintaining the explainability of the model's predictions \cite{explain1,explain12,chen2021adaptive}.

In this study, we present the \textbf{S}patial-\textbf{S}lice \textbf{F}eature \textbf{L}earning \textbf{(SSFL++)} method, an unsupervised approach aimed at reducing computational complexity by effectively eliminating out-of-distribution (OOD) slices and redundant spatial details. Prior studies \cite{hsu2023bag,chen2021adaptive} were not able to pinpoint the most critical slices while also considering the global sequence of information. Recognizing the potential for enhancement in this area, we introduce a \textbf{K}ernel-\textbf{D}ensity-based slice \textbf{S}ampling \textbf{(KDS)} technique, which leverages Kernel Density Estimation to fulfill both objectives concurrently. The results from our experiments showcase the remarkable performance of our 2D model, even in scenarios characterized by a scarcity of data.

Our novelties and contributions can be listed as follow: 
\begin{itemize}

\item \textbf{Improved spatial-slice feature learning module:}
SSFL++ is a morphology-based method tailored for CT scans that efficiently eliminates unnecessary areas in both spatial and slice dimensions. This reduction in computational complexity facilitates the identification of Regions of Interest (RoI) without the need for intricate designs or configurations. Remarkably, we achieved a 70\% reduction in the area on the COVID-19-CT-DB datasets without any decline in performance.


\item \textbf{Density-aware slice sampling method:}
Together with the SSFL++'s capability to selectively eliminate unnecessary spatial areas and slices, the KDS method further refines the process by adaptively identifying and sampling the most essential slices, all while maintaining the integrity of global sequence information. This approach significantly improves data efficiency and bolsters the model's proficiency in few-shot learning scenarios. Experimental outcomes have demonstrated that our E2D model maintains strong and dependable performance, even in situations with limited CT scans and slices.

\end{itemize}
\section{Related Work}
\label{sec:relatedwork}

\subsection{Region of Interests for Computed Tomography}
\textbf{Background.} CT imaging employs X-rays that circulate around a specific body plane, with detectors positioned on the opposite side to capture the emerging signals. This method leverages the varying degrees of X-ray attenuation by different tissues, together with signals collected from numerous angles as they pass through the body, to create a sinogram. This sinogram is instrumental in reconstructing cross-sectional images. However, the intrinsic nature of CT imaging, which requires signal collection from multiple angles for image reconstruction, results in scans filled with superfluous data, potentially leading to increased labor costs in processing and analyzing these images.

Although this technology has been around for a long time, how to design a robust and reliable RoI selection algorithm for CT-scan remains an open-problem. The noises and redundancy harm the model performance. In recent years, most researchers has still focused on how to enhance feature extraction pipeline \cite{ct1}, or improve the quality of image reconstruction \cite{ct2}, to address the aforementioned challenges. Cobo \emph{et al.} \cite{ct3} suggested that the standardization of medical imaging workflows could improve the performance of radiomics and DL systems. Jensen \emph{et al.} \cite{ct4} proposed enhancing the stability of CT radiomics using parametric feature maps. Gaidel \emph{et al.} \cite{GAIDEL2017258} introduced a greedy forward selection-based method for lung CT images, but the lack of robustness against data-shifting and noise limited its development.

Despite the long-standing presence of this technology, crafting a robust and reliable algorithm for selecting Regions of Interest (RoI) in CT scans remains an unresolved issue. The presence of noise and redundant information can adversely affect model performance. In recent years, the focus of most researchers has primarily been on enhancing the feature extraction pipeline \cite{ct1} or improving the quality of image reconstruction to tackle these challenges. Cobo \emph{et al.} \cite{ct3} suggested that standardizing medical imaging workflows could bolster the efficacy of radiomics and deep learning (DL) systems. Jensen \emph{et al.} \cite{ct4} advocated for the stabilization of CT radiomics through the use of parametric feature maps. Meanwhile, Gaidel \emph{et al.} \cite{GAIDEL2017258}introduced a method based on greedy forward selection for lung CT images; however, its lack of robustness against data shifts and noise has hindered its further development.

\subsection{COVID-19 Recognition}

In recent years, significant advancements have been made in developing methods for the detection and recognition of COVID-19 D. Kollias \emph{et al.} \cite{kollias2020deep} have played a pivotal role in this area by delving into how deep learning models can be utilized to analyze prediction outcomes through latent representations. Chen \emph{et al.} \cite{chen2021adaptive} have taken a statistical learning approach, merging maximum likelihood estimation with the Wilcoxon test, to adaptively choose slices and construct models that prioritize explainability. Additionally, Hou \emph{et al.} have explored contrastive learning to refine feature representation, enhancing the model's ability to distinguish between complex patterns. Turnbull \emph{et al.} implemented a 3D ResNet \cite{ResNet} for classifying the severity of COVID-19 cases. Hsu \emph{et al.} introduced a novel two-step model, demonstrating a 2D and (2+1)D strategy \cite{hsu2023bag} that has shown remarkable success in the AI-MIA 2023 competition for COVID-19 detection.

\section{Methodology}

\label{sec:method}

In this section, we present the method we used for COVID-19 detection, including SSFL++ and KDS. The backbone we used is also introduced.

\begin{figure*}
\includegraphics[width=1\textwidth]{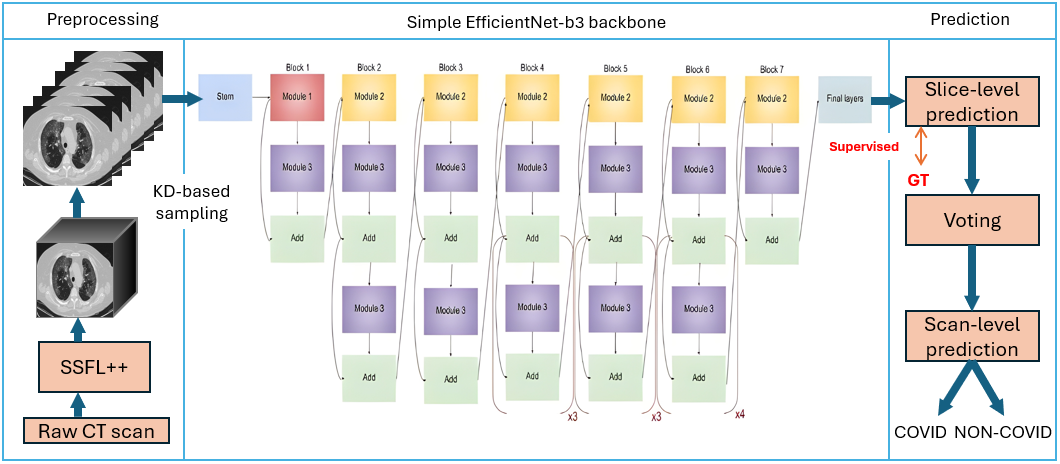}
\caption{Our model architecture (E2D) and pipeline for COVID19-Detection Challenge.} \label{fig:e2d.png}
\end{figure*}

\subsection{Model Architecture}

\textbf{Backbone.} EfficientNet \cite{efficientnet}, through its innovative compound scaling approach, has significantly influenced our perception of achieving a balance between efficiency and accuracy within the realm of deep learning. This methodology, which smartly scales the model's width, depth, and resolution, provides a suite of models that can be efficiently adapted to different hardware limitations.

According to our observations, there is no need to use too complicated or state-of-the art model for recognize COVID-19. Just using ResNet50 \cite{ResNet} or other convolutional neural network like VGG can also archive ideal performance when OOD slices are removed out. In our pipeline, EfficientNet-b3 \cite{rw2019timm} is employed to focus on learning features from individual slices of CT images, utilizing only select portions of these slices. Before moving to the next phase, every slice is processed through EfficientNet to pull out embedded features. This process ensures that only crucial features of the slices are kept within the CT image.

\subsection{Spatial-Slice Feature Learning}


\begin{figure}
\includegraphics[width=0.5\textwidth]{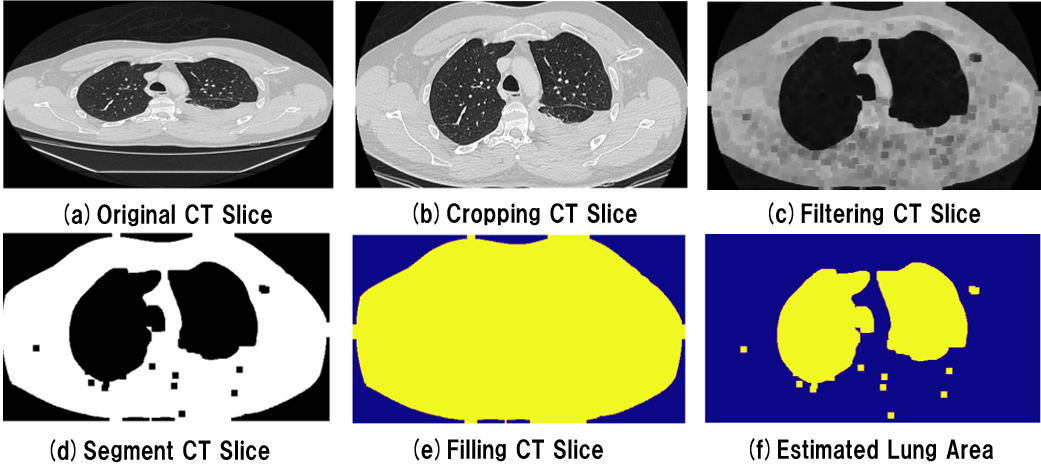}
\caption{The illustration of proposed SSFL++. It aims to remove redundancy both in spatial and slice dimension by simple computed-morphology methods, thereby improving performance and data-efficiency.} \label{fig:spatial.png}
\end{figure}

\textbf{Spatial Steps.} The primary concern with CT scans is the presence of large black areas between the backgrounds of each CT slice. When these images are resized to a fixed shape for input into a neural network, this can distort the RoI, potentially leading to the vanishing of important features. To address this issue, a low-pass filter with a window size of 
$k\times k$ is employed across all CT slices $\mathbf{Z}$ to remove noise. The operation of the low-pass filtering can be defined as follows:

\begin{equation}
\mathbf{Z}_\text{filtered}(i, j) = \frac{\sum_{p=-k}^{k} \sum_{q=-k}^{k} w(p, q) \times \mathbf{Z}(i+p, j+q)}{\sum_{p=-k}^{k} \sum_{q=-k}^{k} w(p, q)}
\end{equation}

where $w(p,q)$ represents the weight at position $(p, q)$ in the filter kernel. The above formula can determine the segmentation $\mathbf{Mask}$ of the filtered slices by a threshold $t$:

\begin{equation}
    \label{eq:seg}
    \mathbf{Mask}[i,j] = 
    \begin{cases}
    0,\,\text{if}\,\mathbf{Z}_\text{filter}[i,j] < t\\
    1,\,\text{if}\,\mathbf{Z}_\text{filter}[i,j] >= t
    \end{cases}
\end{equation}

where i, j denote as an pixel for every single CT slice $\mathbf{Z}^{c}$, which resolution is $x$ $\times$ $y$. A Cropped region $\mathbf{Z}_\text{crop}^{c}$ can be calculated by:

\begin{align*} 
\text{min}(\mathbf{Z}_\text{crop}^{c}(x)) = \min\{i \mid \mathbf{Mask}[i, j] = 1, \forall i\}\\
    \text{max}(\mathbf{Z}_\text{crop}^{c}(x))= \max\{i \mid \mathbf{Mask}[i, j] = 1, \forall i\}\\
    \text{min}(\mathbf{Z}_\text{crop}^{c}(y)) = \min\{j \mid \mathbf{Mask}[i, j] = 1, \forall j\}\\
    \text{max}(\mathbf{Z}_\text{crop}^{c}(y)) = \max\{j \mid \mathbf{Mask}[i, j] = 1, \forall j\}
\end{align*}

$\mathbf{Z}_{crop}^{c}$ is yielded accordingly, we can further resize the resolution of $\mathbf{Z}_\text{crop}^{c}$ to $H$$\times$$W$ for the slice steps and as an input of neural network. Spatial Steps in proposed 4SFL effectively filter out non-lung tissue regions (also known as RoIs in spatial dimension), and reduce computational complexity, as the Figure \ref{fig:spatial.png} illustrated.


\textbf{Slice Steps.} To find the lung tissue region in the CT scan, we used the binary dilation algorithm \cite{enwiki:1082436538} to obtain the filled result $\mathbf{Mask}_\text{filled}$. The difference between the $\mathbf{Mask}$ and filled mask $\mathbf{Mask}_\text{filled}$ represents the lung tissue region. The above method can be summarized as the following formula:

 \begin{equation}
    \label{eq:area}
    Area(\mathbf{Z}) = \sum_i\sum_j\mathbf{Mask}_\text{filled}(i,j) - \mathbf{Mask}(i,j).
\end{equation}

After the above technique, we can finally obtain a range where $s$ and $e$ denote the starting and ending indexes, respectively, and $n_c$ is the constraint of the number of slices for a single CT scan to select most importance RoIs in slice dimension with proportion $\alpha$. The optimization problem can be formulated as following:

Following the application of the low-pass filter technique, we can identify a range where  $s$  and $e$ represent the starting and ending indexes, respectively, and 
$n_c$ denotes the constraint on the number of slices to be selected from a single CT scan. This process aims to select the most important RoIs within the slice dimension based on a proportion $\alpha$. The optimization problem for this selection process can be formulated as follows:

\begin{equation}
    \label{eq:area}
    \begin{split}
    & \underset{s,\,e}{\text{maximize}} \quad \sum^e_{i=s}Area(\mathbf{Z}_i), \\
    & \text{subject to } \quad e-s \leq n_c, \\
    & \quad \frac{\sum^e_{i=s}Area(\mathbf{Z}_i)}{\sum_{i=1}^{n_c} Area(\mathbf{Z}_i)} \geq \alpha.
    \end{split}
\end{equation}



\begin{figure}
\includegraphics[width=0.48\textwidth]{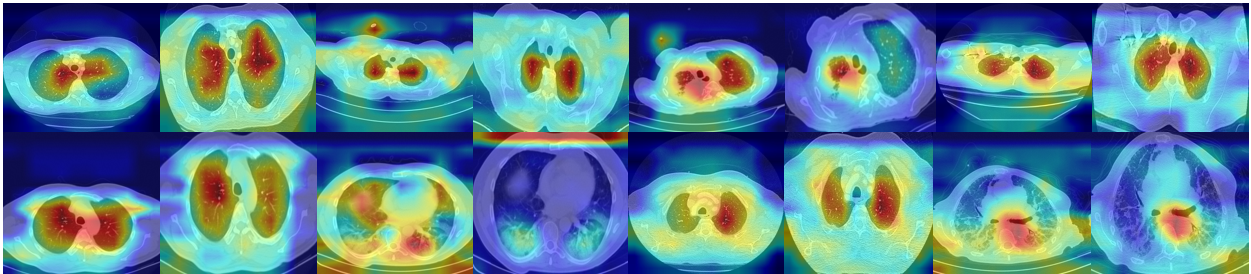}
\caption{The GradCAM++ \cite{gradcam++} visualization before and after proposed SSFL++.} \label{fig:gradcam.png}
\end{figure}

The spatial and slice steps of the proposed SSFL++ model adopt an unsupervised learning approach, relying solely on prior knowledge specific to lung CT scans. This methodology has the potential to be generalized to CT scans of other organs or body parts. However, adapting it to different contexts might necessitate adjustments in parameters to accommodate the unique characteristics of each organ or body part. Furthermore, by employing SSFL++, the visual explanation methods can focus more precisely on RoIs, making them appear more concentrated and thereby facilitating a clearer understanding and analysis of the scans, as shown in Figure \ref{fig:gradcam.png}.

\subsection{Density-aware Slice Sampling}


\textbf{Background.} The SSFL proposed by Hsu \emph{et al.} \cite{hsu2023bag} employs a random sampling method to select slices, which were used for the detection of COVID-19 using 2D and 2+1D CNNs. However, random sampling may potentially introduce bias and instability when training and inference, and it does not efficiently identify the most representative CT slices.

In order to address this, we propose a Kernel-Density-based Slice Sampling (KDS). It performs kernel density estimation (KDE) on the selected slices-set [$\mathbf{Z}_e$,$\mathbf{Z}_s$], adaptively and wisely sampling the most crucial CT-slices. Meanwhile, it also keeps the sequence information globally and alleviates the instability during training and inference stage.

\textbf{Definition.} KDE is a classic method to estimate the probability density function (PDF) of a random variable in a non-parametric manner. It can be defined as:

\begin{equation}
    {\widehat {f}}_{h}(x)={\frac {1}{s}}\sum _{i=1}^{s}K_{h}(x-x_{i})={\frac {1}{sh}}\sum _{i=1}^{s}K\left({\frac {x-x_{i}}{h}}\right)
\end{equation}
\begin{equation}
K(x, x') = \exp\left(-\frac{\|x - x'\|^2}{2\sigma^2}\right)
\end{equation}

where $h$ represents the bandwidth constant, which is determined using Scott's rule \cite{Scott}, $K$ denotes a Gaussian kernel, and $s$ is a smoothing factor applied to the estimated density function (with higher values resulting in smoother functions, set to 100 in this case). Given a Kernel Density Estimation (KDE), several sub-intervals can be established by evaluating its Cumulative Distribution Function (CDF), where the length of each sub-interval adaptively varies according to its $p$-percentile. The CDF of the KDE and its $p$-percentile are calculated using the following formulas:

\begin{equation}
F(x) = \int_{-\infty}^{x} \hat{f}_{h}(t) dt, 
F(q_p) = p
\end{equation}


In the proposed KDS method, the probability of selecting slices within each interval is determined by the density obtained from KDE. Additionally, this method guarantees that at least one sample is selected from each sub-interval, ensuring a comprehensive capture of global sequential information.

\section{Experiment}
\label{sec:experiment}


\begin{table*}[]
    \centering
    \scalebox{1}{
    \begin{tabular}{l|ccc|ccc|ccc}
    \multicolumn{1}{c|}{} & \multicolumn{3}{c|}{Spatial Area (K)} & \multicolumn{3}{c|}{Slice Length} & \multicolumn{2}{c|}{Spatial $\times$ Slice (M)} &  Total \\\cline{2-10}
    \multicolumn{1}{c|}{} & Before & After & $\Delta$ $(\%)$ & Before & After & $\Delta$ $(\%)$ & Before & \multicolumn{1}{c|}{After} & $\Delta$ $(\%)$ \\ \hline
    Training & 267.25 & 155.53 & 0.4184 & 285.32 & 142.91 & 0.4983 & 76.25 & \multicolumn{1}{c|}{22.22} & 0.7085 \\
    Positive & 266.42 & 157.69 & 0.4088 & 295.90 & 148.18 & 0.4985 & 78.83 & \multicolumn{1}{c|}{23.36} & 0.7036 \\
    Negative & 268.21 & 153.03 & 0.4296 & 273.97 & 137.26 & 0.4981 & 73.48 & \multicolumn{1}{c|}{21.00} & 0.7141 \\ \hline
    Validation & 265.62 & 155.23 & 0.4172 & 281.95 & 141.23 & 0.4984 & 74.89 & \multicolumn{1}{c|}{21.92} & 0.7072 \\
    Positive & 268.94 & 160.48 & 0.4061 & 280.53 & 140.55 & 0.4984 & 75.45 & \multicolumn{1}{c|}{22.55} & 0.7010 \\
    Negative & 262.12 & 149.69 & 0.4288 & 283.49 & 141.97 & 0.4984 & 74.30 & \multicolumn{1}{c|}{21.25} & 0.7139 \\ \hline
    (T+V) Positive & 267.25 & 155.53 & 0.4184 & 292.96 & 146.72 & 0.4985 & 78.29 & \multicolumn{1}{c|}{22.81} & 0.7085 \\
    (T+V) Negative & 267.01 & 152.37 & 0.4294 & 275.78 & 138.16 & 0.4982 & 73.64 & \multicolumn{1}{c|}{21.05} & 0.7141 \\ \hline
    Total & 266.94 & 155.47 & 0.4182 & 284.68 & 142.59 & 0.4983 & 75.99 & \multicolumn{1}{c|}{22.16} & 0.7082 \\ \hline
    Testing & 279.55 & 153.41 & 0.4520 & 309.39 & 154.67 & 0.5003 & 86.48 & \multicolumn{1}{c|}{23.72} & 0.7256 
    \end{tabular}
}
\caption{The effectiveness of the SSFL++ module ($k$=$5$,$t$=$100$,$\alpha$=0.5) in eliminating excess data is assessed through three key aspects: spatial, slice, and overall. This method measures how well the SSFL++ module excels at stripping away superfluous details in CT scans, facilitating a more targeted examination and handling. By cutting down on data repetition, the module boosts the speed of computations and could heighten the precision of further analyses or models that utilize the CT data.}
\label{tab:4sfl}
\end{table*}


\textbf{Dataset description.} In our experiments, we used a total of 1,684 COVID-19-CT-DB data, provided by Kollias \emph{et al.} \cite{cvpr24}. The dataset information have shown in Table \ref{tab:dataset_subgroup1}. After a detailed examination, it was noticed that some of the training data included CT scans of the lungs that were not aligned horizontally, either in part or in full. To reduce variability in the training process, these data were manually eliminated. In order to ensure stability and fairly check performance during the experiments, group-5-fold-cross-validation is used. Data augmentation and hyperparameters are kept consistent in all experiments.

\begin{table}[!ht]
    \centering
    \begin{tabular}{c | c c |c} 
       Type& Positive Scan& Negative Scan&   Total Scan \\\hline

      Training     &       703&         655&     1,358\\ 

      Valid     &       170&          156&      326\\\hline
      Total     &       873&          811&      1,684\\\hline
      Testing     &       -&          -&      1,413\\\hline

      Type & Positive Slice & Negative Slice &   Total Slice\\\hline
            Training     &       206,608&         178,722&     385,330\\
      Valid     &       46,042&          43,679&      89,721\\\hline
      Total     &       252,650&          222,401&      475,051\\\hline
      Testing     &       -&          -&      437,185\\
    \end{tabular}
    \caption{The number of data samples at the scan and slice level.}
    \label{tab:dataset_subgroup1}
\end{table}


\textbf{Hyperparameter settings.} The Adam \cite{Adam} optimizer was used with a learning rate of 1e-4 and a weight decay of 5e-4. The batch-size is set to 16.

\textbf{Data Augmentation.} In our experiments, we utilized common augmentation strategy like HorizontalFlip, RandomScaleShifting to prevent overfitting and enlarge feature space. Additionally, we find that HueSaturationValue, RandomBrightnessContrast and CoarseDropout \cite{devries2017improved} are also used. 

\textbf{Loss Function.} The binary cross-entropy loss calculates the cross-entropy between the actual labels and the predicted probabilities, providing an average loss for each sample, where \(y_i\) and \(\hat{y}_i\) represent the actual label and the predicted probability for the \(i^{th}\) sample, respectively.

\begin{equation}
\mathbf{L}_{BCE}(y, \hat{y}) = -\frac{1}{N} \sum_{i=1}^{N} \left[ y_i \log(\hat{y}_i) + (1 - y_i) \log(1 - \hat{y}_i) \right]
\end{equation}

\textbf{Evaluation Metric.} We mainly used F1-score in the experiments for model evaluation. F1-score is a metric used to determine the accuracy of a binary classification model. It combines the harmonic mean of Precision and Recall.

\begin{equation}
\text{f1-score} = 2 \times \frac{\text{precision} \times \text{recall}}{\text{precision} + \text{recall}}
\end{equation}

where precision and recall are computed for COVID and non-COVID. The macro f1-score is the average of the f1-scores for all classes:
\begin{equation}
\text{macro f1-score} = \frac{1}{N} \sum_{i=1}^{N} \text{f1-score}_i
\end{equation}
where $N$ is the number of classes, and $\text{f1-score}_i$ is the f1-score for the $i$-th class. These metrics provide a balanced evaluation of the model's ability to classify each class accurately and its overall performance across all classes.

\subsection{Model Details and Performance Evaluation} To provide a more comprehensive comparison and improve future research, we designed simple E2D in our experiments. The backbones are all based on \textbf{E}fficientNet-b3 \cite{efficientnet,rw2019timm}. The baseline method and detailed pipeline are as follows:

\textbf{Baseline}: The baseline method is presented in \cite{cvpr24}, Kollias \emph{et al.} adopted CNN-RNN to extract feature within all CT-slice. First, all CT-slices are resized to $224$ $\times$ $224$ to extract feature, then RNN (GRU \cite{chung2014empirical} with $128$ neurons) analyzed the 2D-CNN (ResNet-50 \cite{ResNet}) features. The output of the RNN element is then forwarded to a fully connected layer. In addition, this also includes a dropout layer (the dropout rate is set to $0.8$) before the fully connected layer.

\textbf{E2D}: From the CT-scans processed by SSFL++, subsequently, we use our proposed KDS. These sampled slices are resized to $384$ $\times$ $384$ and extracted to high-representation features.



The experimental results, as presented in Table \ref{tab:fewshot}, highlight the E2D model's exceptional performance when paired with KDS on the COVID-19 database 2024 validation set. 


\begin{table}[]
\scalebox{0.82}{
\begin{tabular}{c|cc|c|c}
Model type & Scans& Sampled slice & \begin{tabular}[c]{@{}c@{}}macro f1-score\\ (slice-level)\end{tabular} & \begin{tabular}[c]{@{}c@{}}f1-score\\ (scan-level)\end{tabular} \\ \hline
\begin{tabular}[c]{@{}c@{}}baseline \cite{cvpr24}\end{tabular} & 100\% & - & - & 78.00 \\ \hline

E2D & 100\% & 8(random) & 92.44 & 93.18 \\
 & 100\% & 16(random) & 92.68 & 93.37 \\ 
\multicolumn{1}{l|}{} & 100\% & 8(KDS) & 93.46 & \textbf{100.00} \\
\multicolumn{1}{l|}{} & 100\% & 16(KDS) & \textbf{94.11} & \textbf{100.00}
\end{tabular}
}
\caption{Performance Comparison between baseline provided by Kollias \emph{et al.} \cite{cvpr24}, proposed E2D. We also considered different sampling strategies.}
\label{tab:fewshot}
\end{table}

\subsection{Ablation Study}

\begin{table}[]
\scalebox{0.90}{
\begin{tabular}{ccc|c|c}
Spatial step & Slice step & KDS & \begin{tabular}[c]{@{}c@{}}marco f1-score\\ (slice level)\end{tabular} & \begin{tabular}[c]{@{}c@{}}f1-score\\ (scan level)\end{tabular} \\ \hline
\multicolumn{1}{l}{} & \multicolumn{1}{l}{} & \multicolumn{1}{l|}{} & 80.41 & 81.26 \\
\checkmark &  &  & 88.01 & 88.04 \\
 & \checkmark &  & 90.32 & 90.48 \\
\checkmark & \checkmark &  & 92.68 & 93.37 \\ \hline
\checkmark & \checkmark & \checkmark & \textbf{94.11} & \textbf{100.00}
\end{tabular}}
\caption{The ablation study of proposed SSFL++ and KDS.}
\label{tab:ablation}
\end{table}

To delve deeper into the effects of the SSFL++ and KDS methodologies on the task of detecting COVID-19, an ablation study was carried out, with the findings detailed in Table \ref{tab:ablation}. These experiments all utilize the E2D model, maintaining consistency across experimental hyperparameters. 

The outcomes reveal that the integration of SSFL++ notably boosts performance, highlighting the critical role of spatial redundancy in CT scans and the benefit of precise slice selection. Focusing solely on RoIs proves to be more efficient and effective than analyzing the entirety of CT slices. Conversely, the KDS method further refines the model's predictive capabilities at the slice level and marks substantial advancements at the scan level, demonstrating impressive performance. KDS successfully mitigates the shortcomings of 2D-CNNs, particularly their limited global sequential modeling ability when processing CT images.

\subsection{Submission for COVID-19 Detection Challenge}
\begin{figure}
\includegraphics[width=0.48\textwidth]{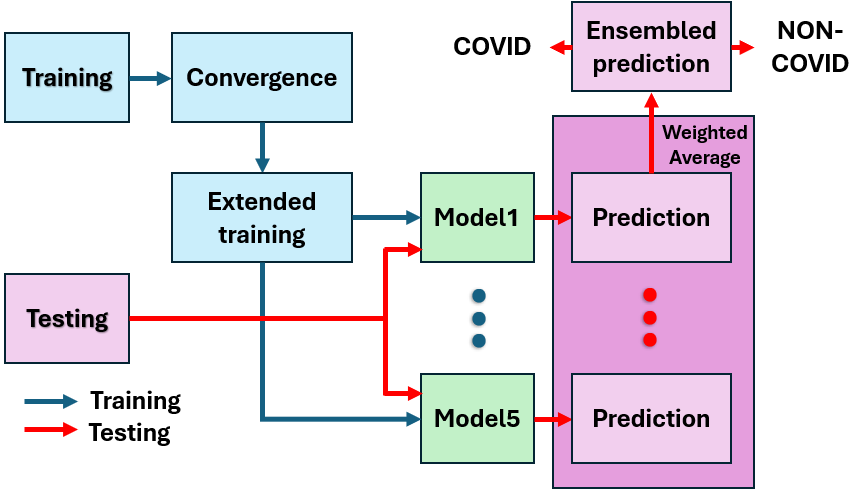}
\caption{The illustration of ensemble strategy for our submission.} \label{fig:ensemble.png}
\end{figure}

We submitted a total of five results. Following the convergence of the neural network, we continued training for several additional epochs and completed model ensembling. Overall, we employed a 5-group validation strategy and used a combination of AUC and slice-level macro F1-score for making predictions. The final output was obtained by averaging these predictive measures. The pipeline is illustrated in Figure \ref{fig:ensemble.png}.






\section{Conclusion}

\label{sec:conclusion}

%

We conducted a comprehensive analysis for the COVID-19 detection task, noting that CT scans often contain a large amount of redundant information, which limits the performance of models. To address this issue, we introduced a simple morphology-based method for CT images, named Spatial-Slice Feature Learning (SSFL++), designed to efficiently and adaptively locate the Region of Interest (RoI). This method effectively reduces redundant areas across both spatial and slice dimensions. 


We combined SSFL++ with the further designed Kernel-Density-based Slice Sampling (KDS), thereby addressing the instability issues brought by random sampling during training and inference phases. Moreover, through the global sequence modeling, we activated the latent capabilities of 2D-CNNs. Finally, our proposed method demonstrated promising results on the validation dataset provided by the DEF-AI-MIA workshop.

{
    \small
    \bibliographystyle{ieeenat_fullname}
    \bibliography{eff}
}

\end{document}